# The $\kappa$ parameter and $\kappa$-distribution in $\kappa$-deformed statistics for the systems in an external field

Guo Lina, Du Jiulin[*]

*Department of Physics, School of Science, Tianjin University, Tianjin 300072, China*

**Abstract**

The $\kappa$-deformed statistics has been studied in many papers. It is naturally important question for us to ask what should the $\kappa$ parameter stand for and under what physical situation should the $\kappa$-deformed statistics be suitable for the statistical description of a system. In this paper, we have derived a formula expression of $\kappa$ parameter based on the $\kappa$-$H$ theorem, the $\kappa$-velocity distribution and the generalized Boltzmann equation in the framework of $\kappa$-deformed statistics. We thus obtain a physical interpretation for the parameter $\kappa \neq 0$ with regard to the temperature gradient and the external force field. We show that, as the *q*-statistics based on Tsallis entropy, the $\kappa$-deformed statistics may also be the candidate one suitable for the statistical description of the systems in external fields when being in the nonequilibrium stationary state, but has different physical characteristics. Namely, the $\kappa$-distribution is found to describe the nonequilibrium stationary state of the system where the external force should be vertical to the temperature gradient.

---

[*] Corresponding author, Email Address: jiulindu@yahoo.com.cn

In recent years, a great deal of attention has been paid to a generalization of Boltzmann-Gibbs (B-G) entropy as well as B-G statistical mechanics so as to deal with complex systems whose properties cannot be exactly described by B-G one. For example, Tsallis $q$-statistics [1], or the called nonextensive statistical mechanics, it has been applied extensively to quite a lot of fields and varieties of interesting problems [2]. In particular, since the connection between the nonextensive parameter $q \neq 1$ and the temperature gradient of the systems in an external potential field is determined by a formula expression [3, 4], it has been known that Tsallis statistics can be reasonably applied to describe the thermodynamic properties of the systems in an external field when they are in the nonequilibrium stationary state, such as stars [5]. And this characteristic for Tsallis nonextensive statistics has received the support recently by the experiment measurements in helioseismology [6].

Basically, in the extended frameworks, the key point lines in the exponential behavior based on the entropy is replaced by a power-law one. Recently, similar motivations underlying generalized statistics also lead to a new example, namely, Kaniadakis entropies [7], such as

$$S_\kappa = -\int d^3 v \left( \frac{z^\kappa}{2\kappa(1+\kappa)} f^{1+\kappa} - \frac{z^{-\kappa}}{2\kappa(1-\kappa)} f^{1-\kappa} \right), \tag{1}$$

where z is the positive real parameter that needs to be specified. Based on this entropy, the $\kappa$-deformed statistics can be given by using so-called $\kappa$-exponential and $\kappa$-logarithm functions defined as

$$\exp_\kappa(f) = \left( \sqrt{1+\kappa^2 f^2} + \kappa f \right)^{1/\kappa}, \tag{3}$$

$$\ln_\kappa(f) = \frac{f^\kappa - f^{-\kappa}}{2\kappa}, \tag{4}$$

where $\kappa$ denotes the deformation parameter. As one may check, we can have $\exp_\kappa(\ln_\kappa f) = \ln_\kappa(\exp_\kappa(f)) = f$ and the above functions could reduce to the standard exponential and logarithm if we let $\kappa \to 0$. In this framework, the generalized $\kappa$-velocity distribution function can be obtained either by maximizing the entropy functional [7] or by the $\kappa$-H theorem in the kinetic theory [8]. It can be expressed as

$$f_\kappa = \frac{1}{z}\exp_\kappa\left(-\frac{mv^2}{2k_B T}\right), \tag{5}$$

where $1/z = (m|\kappa|/\pi kT)^{3/2}(1+3|\kappa|/2)\Gamma(1/2|\kappa|+3/4)/\Gamma(1/2|\kappa|-3/4)$, $k_B$ is Boltzmnn constant and $T$ is temperature. It is clear that the $\kappa$-velocity distribution, Eq.(5), will reduce to Maxwell-Boltzmann (M-B) one if we let the $\kappa$ parameter be zero. Such a $\kappa$-deformed statistics has been studied by Kaniadakis and his co-authors in many papers[9]. It is naturally important question for us to ask what should the $\kappa$ parameter stand for and under what physical situation should the $\kappa$-deformed statistics be suitable for the statistical description of a system. In this letter, we try to give one of the answers.

Let us now consider the system of N particles in the generalized kinetic theory, interacting under the action of an external force field, **F**. The mass of each particle is $m$. We let $f_q(\mathbf{r},\mathbf{v},t)$ be the distribution function of particles in system, and then $f_q(\mathbf{r},\mathbf{v},t)d^3\mathbf{r}d^3\mathbf{v}$ is the particle numbers at time $t$ and in the volume element $d^3\mathbf{r}d^3\mathbf{v}$ around the position **r** and the velocity **v**. And its dynamical behavior is governed by the generalized Boltzmann equation

$$\frac{\partial f_\kappa}{\partial t} + \mathbf{v}\cdot\frac{\partial f_\kappa}{\partial \mathbf{r}} + \frac{\mathbf{F}}{m}\cdot\frac{\partial f_\kappa}{\partial \mathbf{v}} = C_\kappa(f_\kappa), \tag{6}$$

where $C_\kappa$ is called the $\kappa$-collision term. It has been verified by Silva [8] that the solutions of the generalized Boltzmann equation, Eq.(6), satisfy the generalized $\kappa$-H theorem and evolve irreversibly towards the $\kappa$-equilibrium distribution, Eq.(5) (the generalized Maxwellian $\kappa$-velocity distribution). The extension of the $\kappa$-distribution to include the nonuniform systems with interparticle interactions results in

$$f_\kappa = \frac{1}{z}\exp_\kappa\left(-\frac{m(\mathbf{v}-\mathbf{u})^2}{2k_B T}\right), \tag{7}$$

where **u** is the barycentric velocity, and the temperature $T$ should be now a function of the space coordinate **r**. Combining Eq.(3) with Eq.(7) we write

$$f_\kappa = A \left\{ \sqrt{1+\kappa^2 \left[-\frac{m(\mathbf{v}-\mathbf{u})^2}{2k_B T}\right]^2} + \kappa \left[-\frac{m(\mathbf{v}-\mathbf{u})^2}{2k_B T}\right] \right\}^{1/\kappa} \tag{8}$$

where $A=1/z$. The standard M-B distribution is recovered from Eq.(8) if we take $\kappa \to 0$. For convenience, we always let $\mathbf{u} = 0$ by the velocity translation. Then,

$$f_\kappa = A \left\{ \sqrt{1+\kappa^2 \left[-\frac{mv^2}{2k_B T}\right]^2} + \kappa \left[-\frac{mv^2}{2k_B T}\right] \right\}^{1/\kappa}. \tag{9}$$

We now consider the generalized Boltzmann equation. When the $\kappa$-H theorem is satisfied, we have $C_\kappa = 0$ and $\frac{\partial f_\kappa}{\partial t} = 0$ and then Eq.(5) becomes

$$\mathbf{v} \cdot \nabla f_\kappa + \frac{\mathbf{F}}{m} \cdot \nabla_v f_\kappa = 0, \tag{10}$$

where we have used $\nabla = \partial/\partial \mathbf{r}$ and $\nabla_v = \partial/\partial \mathbf{v}$. For the following convenience to use, equivalently we can obtain the equations,

$$\mathbf{v} \cdot \nabla f_\kappa^\kappa + \frac{\mathbf{F}}{m} \cdot \nabla_v f_\kappa^\kappa = 0, \tag{11}$$

and

$$\mathbf{v} \cdot \nabla f_\kappa^{2\kappa} + \frac{\mathbf{F}}{m} \cdot \nabla_v f_\kappa^{2\kappa} = 0, \tag{12}$$

From Eq.(9) we can find out the relation,

$$f_\kappa^{2\kappa} = A^{2\kappa} - 2A^\kappa \frac{\kappa m}{2k_B T} v^2 f_\kappa^\kappa. \tag{13}$$

So,

$$\nabla f_\kappa^{2\kappa} = A^\kappa \nabla A^\kappa - \left(\frac{\nabla A^\kappa}{A^\kappa} - \frac{\nabla T}{T}\right) A^\kappa \frac{\kappa m}{2k_B T} v^2 f_\kappa^\kappa - A^\kappa \frac{\kappa m}{2k_B T} v^2 \nabla f_\kappa^\kappa$$

$$\nabla_v f_\kappa^{2\kappa} = -2 A^\kappa \frac{\kappa m}{2k_B T} \mathbf{v} f_\kappa^\kappa - A^\kappa \frac{\kappa m}{2k_B T} v^2 \nabla_v f_\kappa^\kappa \tag{14}$$

Substituting Eq.(14) into Eq.(12), we have

$$A^\kappa \nabla A^\kappa \cdot \mathbf{v} - \left(\frac{\nabla A^\kappa}{A^\kappa} - \frac{\nabla T}{T}\right) A^\kappa \frac{\kappa m}{2k_B T} \cdot \mathbf{v}^3 f_\kappa^\kappa - 2 A^\kappa \frac{\vec{F}}{m} \frac{\kappa m}{2k_B T} \cdot \mathbf{v} f_\kappa^\kappa$$

$$-A^\kappa \frac{\kappa m}{2k_B T} v^2 \left( \mathbf{v} \cdot f_\kappa^\kappa + \frac{\mathbf{F}}{m} \cdot \nabla_v f_\kappa^\kappa \right) = 0. \tag{15}$$

Then, substituting Eq.(11) and Eq.(9) into Eq.(15), we get the equation,

$$\left[ (\nabla A^\kappa)^2 + A^{2\kappa} \left( \frac{\nabla T}{T} \right)^2 + 2 A^\kappa \nabla A^\kappa \left( -\frac{\nabla T}{T} \right) \right] \left( \frac{\kappa m}{2k_B T} \right)^2 v^4$$

$$+ 4 A^{2\kappa} \left( \frac{\nabla A^\kappa}{A^\kappa} - \frac{\nabla T}{T} \right) \cdot \frac{\mathbf{F}}{m} \left( \frac{\kappa m}{2k_B T} \right)^2 v^2 + 4 A^{2\kappa} \left( \frac{\mathbf{F}}{m} \right)^2 \left( \frac{\kappa m}{2k_B T} \right)^2$$

$$= (\nabla A^\kappa)^2 + 2 \left[ (\nabla A^\kappa)^2 v^2 - A^\kappa \nabla A^\kappa \cdot \frac{\nabla T}{T} v^2 + 2 A^\kappa \nabla A^\kappa \cdot \frac{\mathbf{F}}{m} \right] \left( \frac{\kappa m}{2k_B T} \right)^2 v^2. \tag{16}$$

In this equation, because **r** and **v** are independent variables and Eq.(16) is identically null for any arbitrary **v**, the coefficients of the powers of **v** in Eq.(16) must be zero. Thus, when we consider the coefficient equation for the zeroth-power terms of **v** in Eq.(16), we obtain

$$\left( A^\kappa \frac{\mathbf{F}}{m} \frac{\kappa m}{k_B T} \right)^2 = (\nabla A^\kappa)^2. \tag{17}$$

For the coefficients of the second power of **v**, we find

$$\mathbf{F} \cdot \nabla T = 0. \tag{18}$$

For the coefficients of the fourth power of **v**, we have

$$\left( A^\kappa \frac{\nabla T}{T} \right)^2 = (\nabla A^\kappa)^2. \tag{19}$$

Combining Eq.(17) with Eq.(19), we find

$$\frac{\kappa}{k_B} |\mathbf{F}| = |\nabla T|. \tag{20}$$

In such a way, the $\kappa$-distribution, Eq.(5) or Eq.(9), is shown to describe the thermodynamic properties of the system in an external force field with the characteristics represented by Eq.(18) and Eq.(20). In the relation between the $\kappa$ parameter, the temperature gradient and the external force field, Eq.(20), it is clear that the parameter is $\kappa \neq 0$ if and only if the temperature gradient is $\nabla T \neq 0$. Thus, Eq.(20) presents $\kappa \neq 0$ a clear physical meaning with regard to the temperature gradient and the external force field. If the temperature gradient is $\nabla T = 0$, then we have $\kappa = 0$, which corresponds to the case of B-G statistics, while if the temperature gradient is $\nabla T \neq 0$, then we have $\kappa \neq 0$, which corresponds to the case of $\kappa$-deformed statistics. Therefore, the parameter $\kappa \neq 0$ is shown to be responsible for the spatial inhomogeneity of temperature in the system under the action of an external force field, which corresponds to the nonequilibrium stationary state of such a system.

We need to show that the above relations are derived under the condition of **u** = 0, but they should be still correct in the situation of **u**≠0. Because the velocity **u** is the constant vector, which is not the function of space coordinate **r** and the velocity **v**. While Eq.(16) is identically null for any arbitrary velocity **v**, so it must be the same as for the velocity (**v-u**). Therefore, the results of Eq.(20) and Eq.(18) can be obtained independent on whether **u** is zero or not.

Let us consider the system as the self-gravitating system with the gravitational long-range interparticle interactions in the physical situation of nonequilibrium stationary state, *i.e*, the external force field is now given by $\mathbf{F} = -m\nabla\varphi$, where $\varphi$ is the gravitational potential. In this case, Eq.(18) becomes

$$\nabla\varphi \cdot \nabla T = 0. \tag{21}$$

This relation brings the restriction to bear on the system, where the external force must be vertical to the temperature gradient. And Eq.(20) becomes

$$\kappa m |\nabla\varphi| = k_B |\nabla T|, \text{ or equivalently, } \kappa = k_B |\nabla T| / m |\nabla\varphi|. \tag{22}$$

The form of this relation is found to be very similar to that derived for the nonextensive *q* parameter and the *q*-velocity distribution for the self-gravitating systems [3] and for the nonequilibrium plasma systems [4]. For instance, the parameter $q \neq 1$ for the self-gravitating systems in Tsallis statistics was related to $\nabla T$ and $\nabla\varphi$ by the relation,

$$k_B \nabla T + (1-q) m \nabla\varphi = 0. \tag{23}$$

But it is easy to find out the difference between Eq.(22) and Eq,(23). First, the $\kappa$

parameter must be nonnegative, while (1-*q*) may be positive or negative. Second, from Eq.(21) and Eq.(22) we find that the $\kappa$-distribution can describe the nonequilibrium stationary state of the system, where the external force, $\nabla\varphi$, acts vertical to the temperature gradient, $\nabla T$. But the *q*-velocity distribution in Tsallis statistics describe that physical situation where $\nabla\varphi$ is parallel to $\nabla T$. Therefore, they are both suitable for the statistical description of the systems in an external force field, being in the nonequilibrium stationary states, *but* under different physical situations.

In conclusion, we have derived the formula expression of $\kappa$ parameter based on the $\kappa$-*H* theorem, the $\kappa$-velocity distribution and the generalized Boltzmann equation in the framework of $\kappa$-deformed statistics. We thus obtain a physical interpretation for the parameter $\kappa \neq 0$ with regard to the temperature gradient and the external force field in the system. We show that, as the *q*-statistics based on Tsallis entropy, the $\kappa$-deformed statistics may also be the candidate one suitable for the statistical description of the systems in external fields when being the nonequilibrium stationary state, but has different physical characteristics.

**Acknowledgements**

Guo Lina would like to thank Liu Zhipeng and Qing Fang for their help. This work is supported by the project of "985" program of TJU of China and also by the National Natural Science Foundation of China, No.10675088.

**References**

[1] C.Tsallis, J.Stat.Phys. **52**(1988)479.

[2] M.Gell-Mann, C.Tsallis, 2004, Nonextensive Entropy--Interdisciplinary Applications, Oxford University Press, New York; S.Abe, Y.Okamoto, 2001, Nonextensive Statistical Mechanics and its Applications, Springer-Verlage, Berlin, Heidelberg.

[3] J.L.Du, Europhys.Lett. **67**(2004)893.

[4] J.L.Du, Phys.Lett.A **329**(2004)262.

[5] J.L.Du, New Astronomy **12**(2006)20.

[6] J.L.Du, Europhys.Lett. **75**(2006)861.

[7] G. Kaniadakis, Physica A **296**(2001)405.

[8] R.Silva, Phys.Lett.A **352**(2006)17.

[9] G. Kaniadakis, Phys. Rev. E **66**(2002)056125; G. Kaniadakis and A. M. Scarfone, Physica A, **305**(2002)69, Physica A **340**(2004)102; G. kaniadakis, P. Quariti and A. M. Scarfone, Physica A, **305**(2002)76; M. Cravero, G. Labichino, G. Kaniadakis, E. Miraldi and A. M. Scarfone, Physica A **340**(2004)410; G. Kaniadakis, M. Lissia and A. M. Scarfone, Phys.Rev.E **71**(2005)046128, Physica A **340**(2004)41; S. Abe, G. Kaniadakis and A. M. Scarfone, J. Phys. A **37**(2004) 10513.